\documentclass{webofc}
\usepackage[varg]{txfonts}
\usepackage{paralist}

\begin{document}
\title{GPU-based reconstruction and data compression at ALICE during LHC Run 3}

\author{\firstname{David} \lastname{Rohr on behalf of the ALICE collaboration}\inst{1}\fnsep\thanks{\email{drohr@cern.ch}}
}

\institute{European Organization for Nuclear Research (CERN), Geneva, Switzerland}

\abstract{
In LHC Run 3, ALICE will increase the data taking rate significantly to 50 kHz continuous read out of minimum bias Pb-Pb collisions.
The reconstruction strategy of the online offline computing upgrade foresees a first synchronous online reconstruction stage during data taking enabling detector calibration, and a posterior calibrated asynchronous reconstruction stage.
The significant increase in the data rate poses challenges for online and offline reconstruction as well as for data compression.
Compared to Run 2, the online farm must process 50 times more events per second and achieve a higher data compression factor.
ALICE will rely on GPUs to perform real time processing and data compression of the Time Projection Chamber (TPC) detector in real time, the biggest contributor to the data rate.
With GPUs available in the online farm, we are evaluating their usage also for the full tracking chain during the asynchronous reconstruction for the silicon Inner Tracking System (ITS) and Transition Radiation Detector (TRD).
The software is written in a generic way, such that it can also run on processors on the WLCG with the same reconstruction output.
We give an overview of the status and the current performance of the reconstruction and the data compression implementations on the GPU for the TPC and for the global reconstruction.
}

\maketitle

\section{Introduction}

ALICE (A Large Ion Collider Experiment \cite{bib:alice}) is one of the four major experiments at the LHC (Large Hadron Collider) at CERN.
It is a dedicated heavy-ion experiment studying lead collisions at the LHC at unprecedented energies.
During the second long LHC shutdown in 2019 and 2020, the LHC upgrade will provide a higher Pb--Pb collision rate, and ALICE will update many of its detectors and systems~\cite{bib:aliceupgrade}.
In particular, the main tracking detectors TPC (Time Projection Chamber) and ITS (Inner Tracking System) will be upgraded~\cite{bib:tpcrun3tdr}, and the computing scheme will change with the O$^2$ online-offline computing upgrade~\cite{bib:o2tdr}.

\begin{figure}[htb]
\centering
\includegraphics[width=0.8\textwidth,clip]{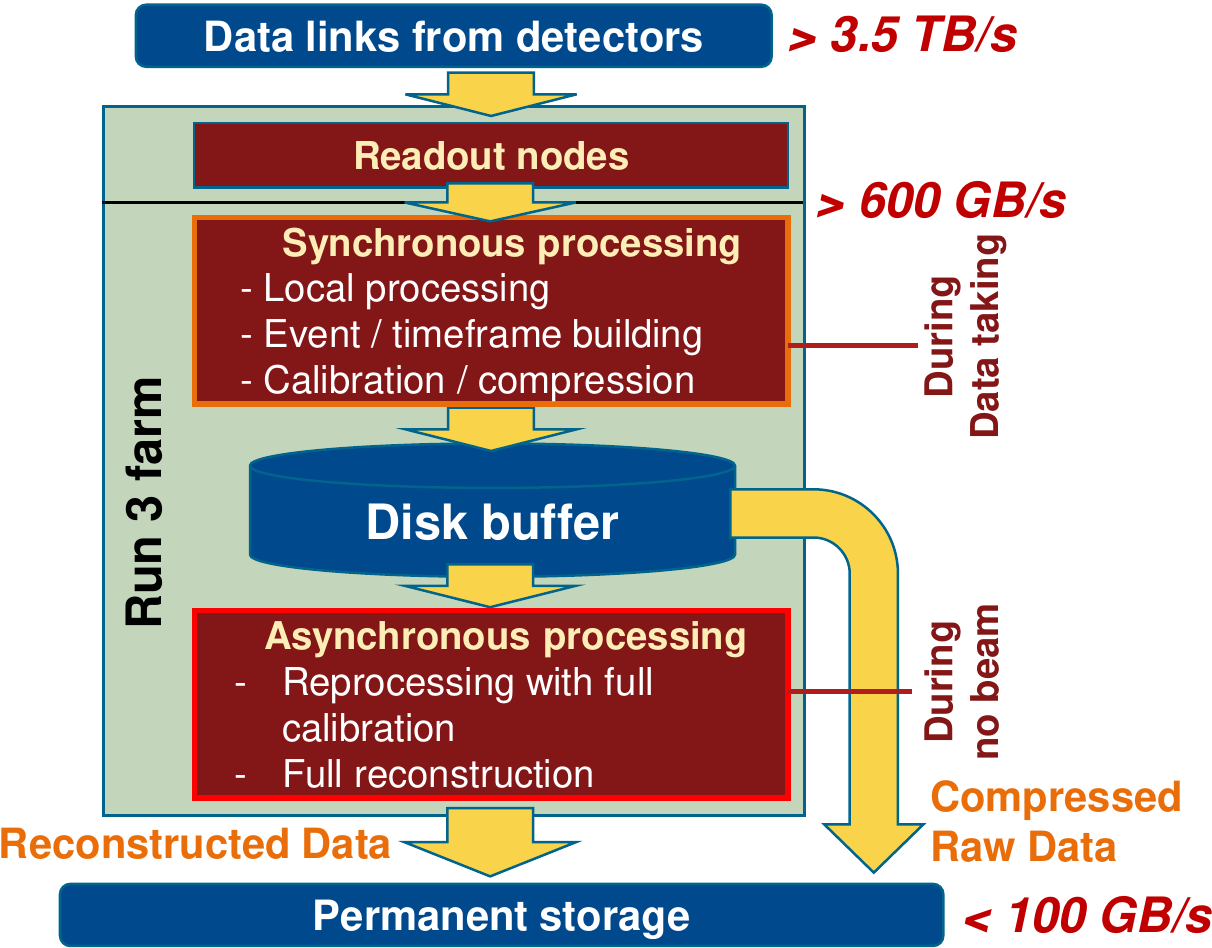}
\caption{Illustration of the ALICE computing strategy for Run 3, with synchronous processing during data taking, and asynchronous processing in periods without beam.}
\label{fig1}
\end{figure}

ALICE will upgrade the detectors for LHC Run~3 and switch from the current triggered read-out of up to 1\,kHz of Pb--Pb events to a continuous read-out of 50\,kHz minimum bias Pb--Pb events.
The continuous read-out of pp collisions will happen at rates between~200\,kHz and~1\,MHz.
ALICE is abandoning the hardware triggers and will switch to a full online processing in software.
During data taking, the synchronous processing will serve two main objectives: detector calibration and data compression.
With a flat budget and the yearly increases of storage capacity, recording and storing raw data as today is prohibitively expensive at 50 to 100 times the data rate.
ALICE aims at a compression of the TPC data, the largest contributor to raw data size, of a factor 20 compared to the zero-suppressed raw data size of Run 2.
By producing the calibration during data taking, ALICE will reduce the number of offline reconstruction passes over the data, where the first two passes serve the calibration today.
The output of the synchronous data processing will be compressed time frames, which are stored to an on-site disk buffer, and from there written to tapes.
When the computing farm is not fully used for the synchronous processing, e.g. in periods without beam or during pp data taking, it will perform a part of the asynchronous reconstruction, which reprocesses the data and generates final reconstruction output.
The part of asynchronous processing that exceeds the capacity of the farm will be done in the grid.
This asynchronous stage will employ the same algorithms and software as the synchronous stage, but with different settings, additional reconstruction steps, and final calibration.
Figure~\ref {fig1} gives an overview of the O$^2$ computing.

\section{GPU Reconstruction for the ALICE Central Barrel Detectors}

\begin{figure}[htb]
\centering
\includegraphics[width=\textwidth,clip]{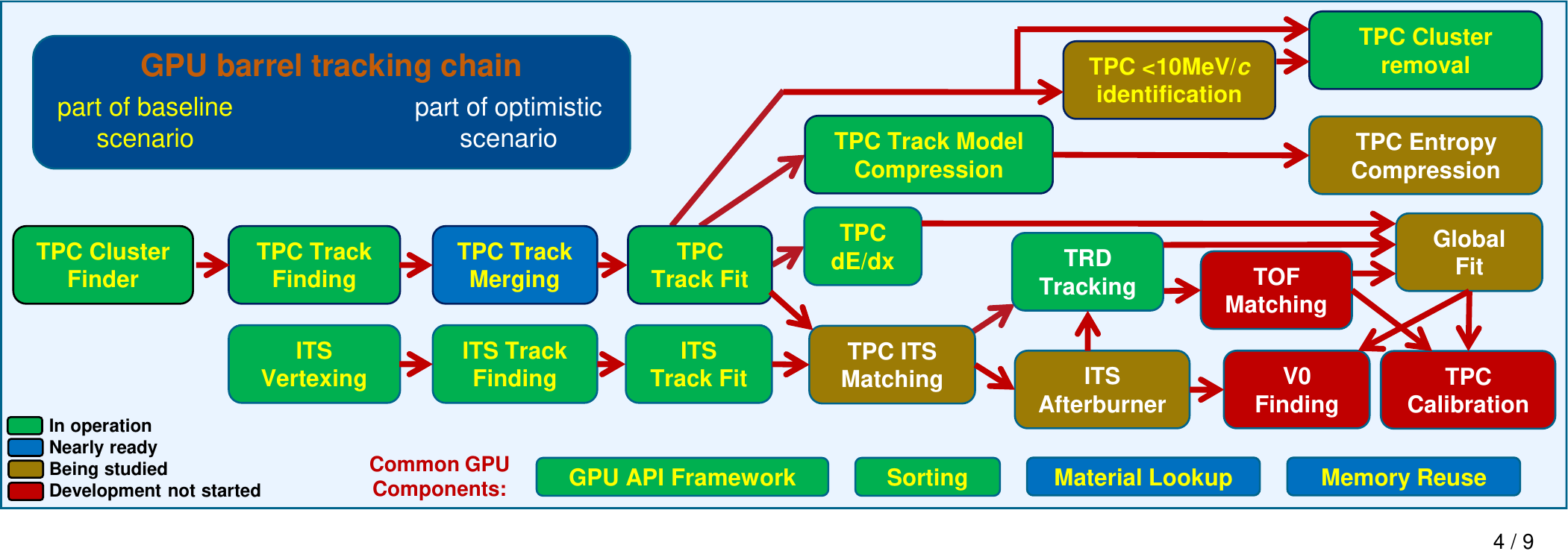}
\caption{Overview of the relevant processing steps in the central barrel tracking and reconstruction chain.}
\label{fig2}
\end{figure}

The reconstruction of the central barrel detectors of ALICE, foremost the TPC (Time Projection Chamber), is the most computing-intense part of event reconstruction, and the focus of this paper.
Therefore, ALICE foresees the usage of Graphics Cards (GPUs) to accelerate these steps.
In parallel, a similar effort on a smaller scale has started to investigate whether the reconstruction of the forward detector reconstruction could leverage GPUs in the same way.
The core part is the tracking of the TPC, which was adapted from the ALICE High Level Trigger~\cite{bib:hltpaper} and improved to match the Run 2 offline reconstruction in terms of efficiency and resolution.
Several new algorithms have been implemented for the GPU reconstruction, in particular for the Inner Tracking System (ITS) \cite{bib:itsgpu}.
Another addition is the data compression for the TPC, which consists of a track model compression step~\cite{bib:lhcp2017} and an entropy encoding step, which will most likely use ANS~\cite{bib:ans} encoding.
We foresee 2 GPU processing scenarios:
\begin{compactitem}
 \item {\bf Baseline scenario}: This contains the minimum set of reconstruction steps on the GPU required to perform the synchronous reconstruction on the online processing farm at the peak data rate assumed for LHC Run 3.
 This scenario defines the size of the online processing farm, in particular the number of processor cores and GPUs.
 \item {\bf Optimistic scenario}: The asynchronous reconstruction will perform many processing steps of the synchronous reconstruction (except for the calibration and the data compression) one more time, thus it can leverage the available GPU algorithms.
 Since there are many more steps in the asynchronous reconstruction, it will be inevitably CPU bound if all these steps are processed by the processor while there are no additional steps on the GPU.
 Therefore we aim to offload more processing steps onto the GPU, and a promising candidate is the complete central barrel tracking chain.
\end{compactitem}
Figure~\ref{fig2} gives an overview of the corresponding reconstruction steps.

We show steps of the baseline scenario with a yellow label, and the additional ones of the optimistic scenario with a white label.
Green boxes indicate steps that are already fully integrated and tested on the GPU, and blue boxes those where a GPU implementation is principally ready but not fully deployed, but there are no significant risks left that could prevent an eventual GPU processing.
For the baseline scenario, most steps are basically ready, except for the TPC Cluster Finder and the identification of TPC tracks below 10 MeV/$c$.
The TPC Cluster Finder is a last-minute addition to the baseline GPU processing.
It was originally designated to run on the FPGA in the readout servers, but it has become likely that the FPGA resources will be insufficient to house the full TPC cluster finding.E-PERF tag.
Therefore, it was moved to the GPU reconstruction.
Consequently, the GPU implementation started late, and it is already in a considerably good shape with only minor features, like the propagation of Monte Carlo labels, missing.
The situation is different for the identification of tracks with low transverse momentum, for which we don't have a working prototype yet that achieves the required efficiency and performance.
As will be discussed in section~\ref{sec:reduction}, it is not clear whether this step will be needed.

The work plan foresees to consolidate the baseline steps first, facilitating a full system test, and then integrating the steps of the optimistic scenario following the order defined by the reconstruction chain graph.
This will make sure that a consecutive set of steps runs on the GPU avoiding unnecessary intermediate data transfer forth and back.
The blocking part for now is the matching of TPC to ITS tracks, which is required for many posterior steps.
The entropy compression, which belongs to the synchronous baseline scenario, is not required to run on the GPU since we already have a sufficiently fast CPU implementation, but it could free up CPU resources for other tasks.
From the estimations of the CPU processing times based on the Run 2 offline reconstruction, the vertex finding represents a considerable CPU load while the tasks seems to be parallel and suited for GPUs.
Therefore, it will make sense to follow the barrel tracking graph up to the V0 finding.

\section{Memory requirements and management}

The ALICE Run 3 processing will be based on time frames, which consist of recorded data over a period of time.
The current design foresees 10 to 20 milliseconds which translates to around 500 to 1000 heavy-ion events at the peak interaction rate of 50 kHz.
Reconstruction of the time frames is performed independently.
This means that tracks from drift detectors like the TPC might range from one time frame to the next one.
Such tracks will not be reconstructible, which will lead to a loss of statistics of less than 1\%, but this simplifies the reconstruction significantly.
However, this means that time frames must be reconstructed as a whole and not be split into parts further.
The TPC is the largest contributor to the data volume.
This means the GPU must either be able to hold the required TPC data for a full time frame, or processing must happen in an approach similar to a ring buffer with the data streamed in and out.
The first approach would be preferable, since the latter makes the software more complicated.
Finally however, a mixture of the two is required since at least the cluster finder will use a ring buffer for its input since the GPU cannot store the full TPC raw data at once.

Many of the steps also use a large scratch memory, which is used temporarily by individual processing steps to store transient results.
These steps must run consecutively on the GPU, and reuse the memory of the previous steps.
Therefore we manage the GPU memory manually.
A large buffer is allocated ahead of time.
Memory is given to certain reconstruction steps, and then reused for the following steps.
For the TPC cluster finding (and also for the TPC track finding~\cite{bib:hltpaper}), the TPC volume is split in 36 sectors which are processed in a pipeline.
The raw data is needed only for the cluster finding step, so once a sector is finished, its raw data con be removed from the GPU.

\begin{figure}[htb]
\centering
\includegraphics[width=\textwidth,clip]{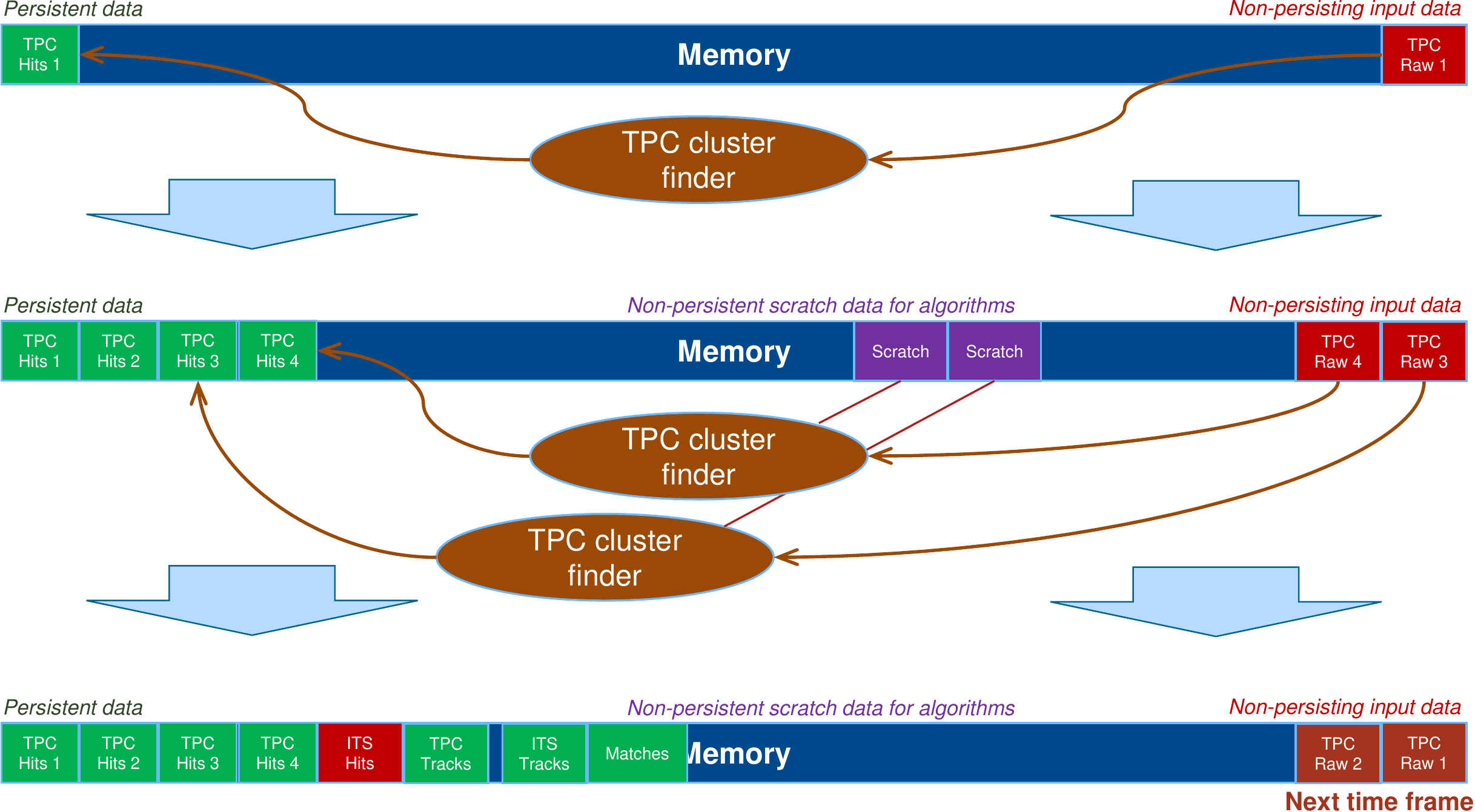}
\caption{Illustration of the memory allocation strategy during the processing of a time frame.}
\label{fig3}
\end{figure}

Figure~\ref{fig3} gives an overview of the memory allocation.
The large buffer is split in a left and a right part.
The left part aggregates data that will persist (e.g. the clusters obtained from the TPC cluster finder, which are used by several subsequent steps).
The right part houses transient data, which is used only by one reconstruction step, and will be overwritten for the next one.
In addition, segments in the middle can be given as scratch buffers temporarily.
The illustration shows from the first to the second row how TPC clusters are aggregated in the persistent region, while the input buffers in the non-persistent region are reused.
Multiple kernels can run in parallel.
They can belong to the same reconstruction step when it runs in a pipeline, or to independent reconstruction steps.
Input data can also be persistent if used many times (like ITS hits), and there may be gaps in the persistent region because in some cases only upper bounds for buffer sizes are available resulting in a gap to the next buffer.
Over time the size of the persistent region increases leaving less scratch space, but this does not create a shortage since the most memory-intense tasks are the TPC cluster finding and track finding, which run at the beginning.
At the end of the processing of one time frame, a special optimization can already preload the first TPC raw data buffers for the next time frame, which minimizes GPU idle time between time frames.

\begin{figure}[htb]
\centering
\includegraphics[width=0.8\textwidth,clip]{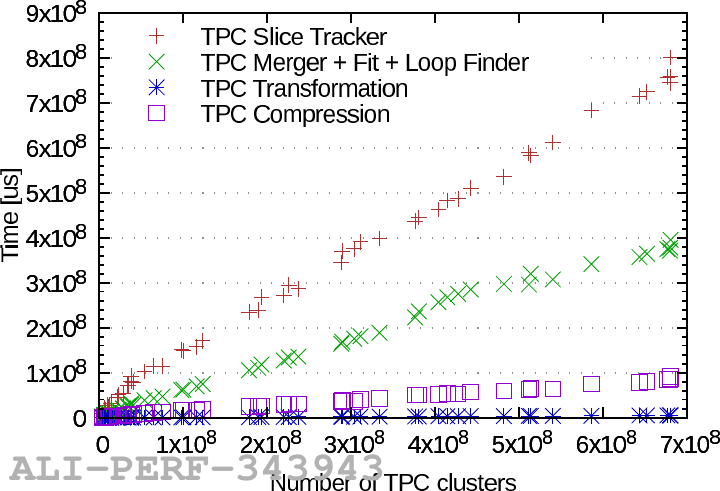}
\caption{Performance of several GPU reconstruction steps versus size of input data.}
\label{fig4}
\end{figure}

Figure~\ref{fig4} shows that the processing time of the individual steps depends basically linearly on the input data size and thus on the length of the time frame with only minor fluctuations.
This is important since it proves that the approach of processing long time frames at once does not produce an overhead in processing time.

With the current implementation, the ALICE reconstruction will need 16 GB of memory for the processing of a 10 ms time frame.
Processing longer time frames will require GPUs with more memory, which will probably be much more expensive.
While some optimizations are possible, it is not clear whether the memory requirements can be reduced sufficiently to operate on GPUs with less than 16 GB of memory without a ring buffer.

\section{TPC Cluster Removal}

\label{sec:reduction}

The compression of the TPC data involves multiple steps:
\begin{compactitem}
\item Clusterization of the raw data into clusters (lossy) \cite{bib:hltpaper}.
\item Track model and entropy compression (lossless) \cite{bib:hltpaper, bib:lhcp2017}.
\item Removal of clusters of tracks that will not be used for physics analysis (lossy) \cite{bib:ctd2019}.
\end{compactitem}
The third point, removal of clusters, can be realized in two ways:
\begin{compactitem}
 \item {\bf Strategy A}: Positive identification of clusters and tracks to be removed.
 \item {\bf Strategy B}: Identification of good tracks to be kept, and removal of everything else.
\end{compactitem}
In both cases, the clusters in a tube around the good tracks are protected from removal to ensure the optimal tracking resolution with final calibration which can modify the cluster attachment.
Strategy B will be faster and remove more clusters than Strategy A, but it bears the risk of removing clusters of good tracks if the synchronous reconstruction was unable to reconstruct a good track, or reconstructed it incompletely.
Since the tracking algorithm is the same, the difference will depend to the largest extent on the calibration.
Currently, both strategies are developed in parallel until the implications of strategy B are fully understood.
So far, strategy A lacks the identification of tracks below 10 MeV/$c$ as shown in Fig.~\ref{fig2}, which reduces the achievable reduction factors significantly.
Currently, strategy A yields a final data rate of 87.7 to 118.1 GB/s of compressed raw data transferred to the storage for 50 kHz Pb--Pb, while strategy B achieves 71.7 to 89.9 GB/s.
The O$^2$  Technical Design Report (TDR) assumes an output rate of 88 GB/s.
In these ranges, the higher bound represents the current state of the software, while the lower bound uses Monte Carlo information to estimate the highest achievable reduction if track merging and protection of clusters in the tube around good tracks were 100\% efficient.
 
\section{Conclusions}

We have presented the online computing strategy of ALICE for Run 3 and the central barrel tracking chain, which is the most promising candidate for GPU usage.
The implementation of the software for the baseline scenario, which will run the computing-intense synchronous reconstruction steps on GPUs, is nearly finished while the offloading of additional steps for the optimistic scenario is ongoing.
By reusing the same memory for consecutive processing steps, GPUs with 16 GB of memory can process time frames of around 10 ms.
Longer time frames will require more GPU memory, and the usage of GPUs with less memory would necessitate significant software changes and the inclusion of a ring buffer.
TPC data reduction strategy B yields the foreseen data rates of the TDR today but bears the risk of losing good tracks.
Strategy A does not bear this risk but does not yet achieve the desired data reduction factors.
Its implementation and optimization are ongoing.

\end{document}